\documentclass{article}\usepackage{natbib,emulateapj}
\usepackage{psfig}
\newcommand{\beq}{\begin{equation}}
\newcommand{\eeq}{\end{equation}}
\newcommand{\bea}{\begin{eqnarray}}
\newcommand{\eea}{\end{eqnarray}}
\newcommand{\bean}{\begin{eqnarray*}}
\newcommand{\eean}{\end{eqnarray*}}
\newcommand{\ba}{\begin{array}}
\newcommand{\ea}{\end{array}}
\newcommand{\bml}{\begin{mathletters}}
\newcommand{\eml}{\end{mathletters}}
\newcommand{\rem}[1]{{ }}
\def\r2{\langle r^2 \rangle}
\def\lpm{l_{\perp,{\rm min}}}
\bibpunct[,]{(}{)}{;}{a}{}{,}
\begin{document}
\title{Thermal Conduction in Clusters of Galaxies} 
\author {Ramesh Narayan\altaffilmark{a} and 
Mikhail V. Medvedev\altaffilmark{b} } 
\altaffiltext{1}{Institute for Advanced Study, Princeton, NJ 08540;
Permanent address: Harvard-Smithsonian Center for
Astrophysics, Cambridge, MA 02138; rnarayan@cfa.harvard.edu}
\altaffiltext{2}{Canadian Institute for Theoretical Astrophysics,
Toronto, ON, M5S~3H8, Canada; medvedev@cita.utoronto.ca;
http://www.cita.utoronto.ca/$\sim$medvedev/}

\begin{abstract}
We estimate the thermal conductivity of a weakly collisional
magnetized plasma with chaotic magnetic field fluctuations.  When the
fluctuation spectrum extends over two or more decades in wave-vector,
we find that thermal conduction is very efficient; the conduction
coefficient is only a factor $\sim5$ below the classical Spitzer
estimate.  We suggest that conduction could play a significant role in
cooling flows in clusters of galaxies.
\end{abstract}

\section{Introduction}

Hot X-ray-emitting gas is ubiquitous in clusters of galaxies
\citep{Sarazin88}. Since the X-ray emission is energetically important
in many clusters, it is believed that a significant amount of mass
must continuously cool and drop out of the intracluster medium
\citep{Fabian94}.  The mass deposition rate is estimated to be as much
as several hundred $M_\odot\,{\rm yr^{-1}}$ in some clusters (e.g.,
\citealp{David+01,Allen+01}).

Direct evidence for the cooling gas has, however, been scarce
\citep{Fabian94}. In particular, recent observations with {\it
XMM-Newton} \citep{Boeh01,MP01} and {\it Chandra} \citep{F2001} have failed
to find the multi-temperature gas one expects in a cooling flow.  The
observations suggest that mass dropout may be less significant than
previously thought.  A reduced level of mass dropout is possible if
there is a source of heat to replace the energy that is lost through
X-ray emission, but no clear heat source has yet been identified
\citep{F2001}.

The inner region of a cluster ($R\sim$ few $\times10$ kpc) where mass
dropout appears to be occurring is typically cooler than the rest of
the cluster.  Therefore, an often-discussed source of heat is thermal
conduction from the hot outer regions of the cluster to the center
\citep{BC81,TR83,BM86,BD88,G89,RT89,PS96,S01}. While the idea is
attractive, it requires extremely efficient conduction, which is
considered problematic.

In a classic paper, \citet{Spitzer62} showed that thermal conduction
in an unmagnetized plasma has a diffusion constant\footnote{ The
coefficient $\kappa_c$ which enters in the heat flux equation,
$q=-\kappa_c\nabla T$, is related to $\kappa_{\rm Sp}$ by $\kappa_c=n
k_B\kappa_{\rm Sp}$, where $k_B$ is the Boltzmann constant.  Note also
that heat diffusion and particle diffusion have slightly different
coefficients, differing by a factor of order unity.}  $ \kappa_{\rm
Sp}\sim\lambda^2/t_{\rm Coul} =\lambda v_t\sim4\times10^{32}\;
T_1^{5/2}\,n_{-3}^{-1} \textrm{~~cm}^2\textrm{~s}^{-1}, $ where
$t_{\rm Coul}=\lambda/v_t$ is the mean free time between Coulomb
collisions, and $\lambda$ and $v_t$ are the mean free path and the
thermal speed of electrons (\citealp{CMc77,EF00}): $ \lambda\sim30\;
T_1^2\, n_{-3}^{-1}\textrm{~~kpc}, ~ v_t\sim(kT/ m_e)^{1/2}\sim
4\times10^9\; T_1^{1/2} \textrm{~~cm~s}^{-1}$.  Here, $T_1=kT/10~{\rm
keV}$ is the scaled temperature, $n_{-3}=n/10^{-3}~{\rm cm}^{-3}$ is
the scaled electron number density, and we have used an average value
for the Coulomb logarithm, $\ln\Lambda\sim38$.

The time required for heat to diffuse conductively across a radius
$R=100R_2$ kpc is given by $ t_{\rm Sp}\sim R^2/\kappa_{\rm
Sp}\sim8\times10^6 \,T_1^{-5/2}n_{-3}R_2^2 ~~{\rm yr}$.  For
conduction to have a significant effect on a cooling flow, the
conduction time must be comparable to the cooling time $t_{\rm cool}$
of the gas.  Table 1 lists representative data for two clusters,
Hydra~A \citep{David+01} and 3C295 \citep{Allen+01}, at two
characteristic radii, 100 kpc and 10 kpc.  Columns 6 and 7 give
$t_{\rm cool}$ and $t_{\rm Sp}$.  We see that if thermal conduction in
a cluster is as efficient as in Spitzer's theory, or even if it is a
factor of a few less efficient, heat conduction will have a strong
effect on the energetics of a cooling flow and perhaps will shut off
mass dropout.  The main problem with this idea is that the gas in a
cluster is likely to be magnetized, and conventional wisdom says that
magnetic fields severely suppress conduction relative to the Spitzer
level.  This is the topic of the present {\it Letter}.

We discuss in \S2.1 the theory of conduction in a tangled magnetic
field as developed by Rechester \& Rosenbluth (1978, RR) and Chandran
\& Cowley (1998, CC); the theory predicts that the coefficient of
thermal conduction is a factor $\sim100-1000$ lower than the Spitzer
coefficient.  We then present in \S2.2 and \S2.3 an extension of the
theory to a turbulent medium; we show that if turbulence extends over
a factor of 100 or more in length scale, thermal conduction is almost
as efficient as in Spitzer's theory.  We conclude with a brief
discussion in \S3.

\section{Theory of Thermal Conduction in a Weakly Collisional Magnetized Gas}

\subsection{Conduction in a Chaotic Magnetic Field With a Single Scale}
\label{S:RR}

In the presence of an ordered magnetic field, conduction is
anisotropic.  Electrons stream freely parallel to the field line, so
the parallel diffusion constant is almost equal to the Spitzer value:
$\kappa_\parallel\sim\kappa_{\rm Sp}/3$.  (The factor of $1/3$ is
because diffusion is in one dimension rather than three, see CC).
Perpendicular to the field, however, electrons follow circular Larmor
orbits with radius $\rho_e\ll \lambda$.  Since an electron moves only
a distance $\sim\rho_e$ in each scattering, the perpendicular
diffusion constant is given by $ \kappa_\perp\sim\rho_e^2/t_{\rm
Coul}\sim(\rho_e/\lambda)^2\kappa_{\rm Sp} \ll\kappa_{\rm Sp}$.  For a
galaxy cluster with a magnetic field of $\sim10^{-6}$ G, we have
$\rho_e\sim10^{-12}\lambda$, and $\kappa_\perp$ is effectively zero.

Thermal conduction behaves very differently when the magnetic field is
chaotic.  The theory for a tangled field with a single coherence
length $l_B$ was developed by RR and has been recently revived in the
astrophysical context by CC.  Since the field is chaotic, the
separation $r$ of two nearby field lines must have a Lyapunov-like
scaling as a function of distance $l$ along the field: \beq r \sim
r_0\,\exp(l/L_{\rm Lyap}),
\label{d(l)}
\eeq where $r_0$ is the initial separation of the two lines.  As there
is only one characteristic scale in the problem, namely $l_B$, we
expect $L_{\rm Lyap}\sim l_B$.

Following RR and CC, let us consider the evolution of a compact cloud
of electrons of initial size $\rho_e$.  With time, the electrons
diffuse parallel to the field, with a diffusion constant
$\kappa_\parallel$.  As the electron cloud spreads out, its
perpendicular extent diverges exponentially according to equation
(\ref{d(l)}).  Thus, by the time the electrons have diffused a
Rechester-Rosenbluth distance $ L_{RR}\sim l_B\ln(l_B/\rho_e)\sim
30l_B$ along the field line, their transverse separation is of order
$l_B$.  The numerical coefficient 30 corresponds to
$l_B/\rho_e\sim10^{13}$, a typical value for a galaxy cluster
(assuming $l_B$ is a fraction of the radius).  Being a logarithmic
factor, the numerical value is insensitive to details.  When electrons
have moved a distance $L_{RR}$ along the tangled field line, their
three-dimensional root mean square (rms) displacement is $R_*$, where
$ R_*^2\sim L_{RR}l_B\sim30l_B^2. $ Beyond $R_*$, the motion of an
electron is isotropic and uncorrelated with its previous path.

Let us define $t_*$ as the time it takes for electrons to diffuse a
distance $L_{RR}$ along the field: $
t_*\sim{L_{RR}^2/\kappa_\parallel}.  $ For $t<t_*$, electrons diffuse
anisotropically: $l\sim(\kappa_\parallel t)^{1/2}, 
r\sim\rho_e\exp[(\kappa_\parallel t)^{1/2}/l_B]$.  For $t>t_*$,
however, electrons diffuse isotropically, and move in three dimensions
according to $ R\sim(\kappa_*t)^{1/2}$, where
\beq\kappa_*\sim{R_*^2/t_*} \sim(l_B/L_{RR})\,\kappa_\|
\sim10^{-2}\kappa_{\rm Sp}.\label{kappa*} \eeq

We see that for $R>R_*$ the conduction is many orders of magnitude
more efficient than when the field is ordered.  However, $\kappa_*$ is
still a factor $\sim100$ less than $\kappa_{\rm Sp}$.  The conduction
time is correspondingly $\sim100$ times longer than the Spitzer time
$t_{\rm Sp}$.  As Table 1 shows, such weak conduction is unlikely to
have an important effect on cooling flows.

The estimate given in equation (\ref{kappa*}) is valid so long as
$\lambda<l_B$.  This condition, which is likely to be satisfied by the
gas in clusters (compare $\lambda$ with $l_B\sim R/({\rm few})$ in
Table 1), ensures that collisions enable electrons to pass through
magnetic mirrors caused by inhomogeneities in the field.  If
$\lambda>l_B$, conduction is suppressed by an additional factor
$\theta<1$ (\citealp{C99,MK01}) since only a fraction of the electrons
are able to penetrate the mirrors.  This would cause the conduction
time to increase by a factor $1/\theta$.

\subsection{Conduction in a Multiscale Chaotic Magnetic Field}

A key assumption of the RR theory is the presence of a single Lyapunov
length scale $L_{\rm Lyap}\sim l_B$.  However, if the medium is
turbulent, chaotic fluctuations will be present over a wide range of
length scales.  We generalize the theory for such a multiscale medium.

We begin by re-expressing the single-scale theory as follows.  When
two field lines are separated by a distance $r$ smaller than $l_B$,
their mean square separation $\r2$ increases with distance $l$ along
the field line according to a Lyapunov-like scaling.  However, when
$r>l_B$, the increase is given by the usual diffusion law, where $\r2$
increases by $\Delta r^2\sim l_B^2$ for a parallel displacement
$\Delta l\sim l_B$.  Thus, we may describe the evolution in the two
regimes, $\r2<l_B^2$ and $\r2>l_B^2$, by the following two
differential equations:\beq {d\r2 / dl} \sim 2{\r2 / l_B}, \qquad
{d\r2 / dl} \sim {l_B^2 / l_B} = l_B.
\label{RRdiff} \eeq

Consider now a tangled magnetic field with a range of scales, and
assume that the statistics of the magnetic field fluctuations are
described by the Goldreich \& Sridhar (1995, GS) theory of Alfvenic
MHD turbulence.  In a GS turbulent cascade, there is a range of scales
$l_\perp$ perpendicular to the field, extending from a minimum scale
$\lpm$ to a maximum scale $l_B$.  The fluctuations are anisotropic, so
that for a given perpendicular scale $l_\perp$ the corresponding
parallel coherence scale $l_\parallel$ is given by \beq (l_\|/l_B)\sim
(l_\perp/l_B)^{\alpha}, \quad \lpm<l_\perp<l_B. \label{aniso} \eeq For
simplicity, we have selected the normalization such that on the outer
scale $l_B$, the fluctuations are isotropic: $l_\perp\sim
l_\parallel\sim l_B$.  The index $\alpha$ is equal to 2/3 for strong
MHD turbulence (GS) and 3/4 for intermediate turbulence
(\citealp{GS97}).

Using the single-scale equations (\ref{RRdiff}) as a guide, it is
straightforward to write down corresponding equations for a medium
with a spectrum of fluctuations.  We then identify three regimes for
the evolution of $\r2$.

First, when $r\equiv\r2^{1/2}<\lpm$, all the fluctuation scales in the
medium contribute to Lyapunov-like growth (assuming they all behave
chaotically), so \beq {d\r2 \over dl} \sim 2\r2\int_{1/l_B}^{1/\lpm}
{d\ln k_\perp \over l_\parallel} \sim {2\r2\over l_{\parallel,{\rm
min}}},
\label{regime1}\eeq where we have written
$k_\perp=1/l_\perp$, and we have ignored constants of order unity in
the normalization of the integral.  We see that the effective Lyapunov
scale for the growth of $r$ is the parallel coherence length of the
{\it smallest} scale fluctuations in the medium, i.e.  \beq r\sim
r_0\,\exp(l/l_{\parallel,{\rm min}}), \qquad l_{\parallel,{\rm min}}
\sim \lpm^{\alpha} l_B^{1-\alpha}.
\label{soln1}
\eeq Since usually $l_{\parallel,{\rm min}}\ll l_B$, the growth is
rapid.

Once $r$ exceeds $\lpm$, the evolution switches to a second regime.
We continue to have Lyapunov-like growth from scales $l_\perp>r$, but
there is diffusion-like growth for scales $l_\perp<r$.  The evolution
equation for $\lpm^2 < \langle r^2\rangle < l_B^2$ thus becomes \beq
{d\r2 \over dl} \sim 2\r2\int_{1/l_B}^{1/r} {d\ln k_\perp \over
l_\parallel} + \int_{1/r}^{1/\lpm}d\ln k_\perp {l_\perp^2 \over
l_\parallel}.
\label{main}
\eeq If $l_B\gg r\gg\lpm$, each integral in equation (\ref{main}) is
dominated by the scale $r$.  Substituting equation (\ref{aniso}) in
equation (\ref{main}), we then obtain the solution \beq
r/l_B\sim(l/l_B)^{1/\alpha}.
\label{soln2}
\eeq Remarkably, the separation between two neighboring field lines
becomes of order $l_B$ for a parallel translation of only $\sim l_B$;
this is much faster than in the RR theory which requires a parallel
translation $\sim30l_B$.  The solution $r\propto l^{1/\alpha}$
corresponds exactly to $l_\perp\propto l_\|^{1/\alpha}$ in the
turbulence model (\ref{aniso}).  Thus, the rms separation of field
lines grows along the ``Goldreich-Sridhar cone.''

When $r>l_B$, we enter a third regime, which corresponds to isotropic
diffusion.

>From equation (\ref{soln2}), it is clear that thermal conduction in a
multiscale chaotic field is almost as efficient as in Spitzer's
theory.  Replacing $L_{RR}$ by $l_B$ in equation (\ref{kappa*}), we
estimate $\kappa_{\rm turb}\sim \kappa_{\rm Sp}/3$ (but see \S2.3 for
a better estimate of the coefficient).  As in the previous subsection,
we have assumed $\lambda<l_B$ and have not included a magnetic mirror
factor $\theta$.  In GS turbulence, perturbations on length scales
$l_\perp<l_B$ have weak magnetic field fluctuations, $\Delta B/B\sim
l_\perp/l_\parallel\sim (l_\perp/l_B)^{1-\alpha}<1$, and cause
negligible mirroring.  Only perturbations on the scale $l_B$ cause
strong mirroring, but these have a negligible effect so long as
$\lambda<l_B$ \citep{C99,MK01}.

\subsection{Numerical Solutions}

We have numerically integrated the differential equations
(\ref{regime1}) and (\ref{main}), starting with an initial separation
$r=10^{-13}l_B\sim\rho_e$, and assuming $\alpha=2/3$ as appropriate
for strong turbulence in the GS model.  Figure 1 shows four numerical
solutions for the evolution of $r$ as a function of distance $l$ along
the field line, corresponding to four choices of the minimum scale
$\lpm$ of the turbulence.  We see exponential growth of $r$ for
$l<\lpm$, and power-law growth for larger separations, confirming the
scalings given in (\ref{soln1}) and (\ref{soln2}).

Let us define the decorrelation length $L_{\rm dec}$ as the distance
along the field line for which the transverse separation $r$ becomes
equal to $l_B$.  Figure 2 shows how $L_{\rm dec}$ depends on $\lpm$.
When $\lpm\sim l_B$, the turbulence is dominated by a single (outer)
scale.  This corresponds to the RR theory, and in this limit the
decorrelation length is large, as expected.  However, for $\lpm\la
10^{-2}l_B$, we find that $L_{\rm dec}$ is quite small, asymptoting to
$\sim1.6l_B$.  Since $L_{\rm dec}$ is the analog of the
Rechester-Rosenbluth length $L_{RR}$ for a multiscale medium, we may
replace $L_{RR}$ by $L_{\rm dec}$ in equation (\ref{kappa*}) to
estimate the diffusion constant in a turbulent medium\footnote{Since
eqs. (\ref{regime1}), (\ref{main}) are approximate and may contain
numerical coefficients of order unity multiplying the integrals, the
numerical factor of 1/5 in eq. (\ref{kappa-turb}) is approximate as
well. } \beq \kappa_{\rm turb}\sim (l_B/L_{\rm dec})\,
\kappa_\parallel \sim\kappa_{\rm Sp}/5, \qquad t_{\rm turb}\sim5t_{\rm
Sp}.
\label{kappa-turb}
\eeq Thus, if turbulence extends over at least two decades in scale,
conduction is very efficient and approaches the Spitzer level to
within a factor of a few.

\section{Discussion}
\label{S:DISC}

Thermal conductivity in a homogeneous magnetic field is known to be
highly anisotropic --- it is Spitzer along the field, but
extraordinarily reduced in the transverse direction.
RR came up with the important insight that when the magnetic field is
tangled and chaotic, thermal conduction is enhanced significantly by
the exponential divergence of neighboring field lines.  However, even
with this effect, CC estimated that the conductivity in galaxy
clusters is below Spitzer by a factor $\ga100$.

We have shown in this {\it Letter} that if the field is chaotic over a
wide range of length scales (factor of 100 or more), as might happen
with MHD turbulence (GS), thermal conduction is boosted to within a
factor $\sim5$ of the Spitzer value.  Such strong conduction will have
a significant effect on galaxy clusters (compare columns 6 and 8 in
Table 1).  It can transport heat to the center of a cluster to replace
the energy lost through cooling, and it can also eliminate any thermal
instability in the cooling gas.  Thus, it may well reduce the need for
large-scale mass dropout in cooling flows \citep{TR83,BM86}.  Only in
the inner regions of some clusters (e.g. 3C295, Table 1) might there
be significant dropout.  It is worth noting that some authors have
discussed potential problems with invoking such strong conduction
\citep{BC81,BD88}, while others have rebutted these arguments
(\citealp{RT89,S01}).

An important requirement for the validity of our analysis is that the
magnetic field should behave chaotically, i.e., it should exhibit
Lyapunov-like behavior over a wide range of scales.  Weak MHD
turbulence consists of a superposition of Alfv\'en waves and is not
chaotic.  However, the model of strong and intermediate MHD turbulence
developed by GS is chaotic, as indicated by the breakdown of
perturbation theory (Goldreich \& Sridhar 1997).  

It should be noted that our theory of thermal conduction does not
require ongoing dynamic turbulence.  Each episode of dynamic
turbulence will leave behind a substantial level of frozen-in tangled
fields even after the turbulent motions have ceased.  Such leftover
tangled fields should be sufficient to enhance conduction to the
levels we estimate.

As a final point, we should discuss a serious caveat.  Chandra has
found evidence for sharp temperature jumps in a few clusters, e.g.,
Abell 2142 \citep{M+00} and Abell 3667 \citep{V+01}.  The observations
indicate that conduction across the temperature jumps is far below
Spitzer \citep{M+00,EF00}, in apparent contradiction with the estimate
presented here.  \citet{V+01} propose that the magnetic field is
stretched parallel to the interface, thus inhibiting diffusion across
the field on small scales.  Extending this idea, we suggest that when
two distinct objects merge, as appears to be the case with the above
clusters, the two regions (each of which is internally chaotic and
highly conducting) may be thermally isolated from each other because
their magnetic fields have not yet interpenetrated each other.  It is
unclear how long such magnetic isolation will survive.

\acknowledgements 

The authors thank Stas Boldyrev and Andrei Gruzinov for useful
discussions and the referee for helpful comments.  RN was supported in
part by the W.M. Keck Foundation as a Keck Visiting Professor at the
Institute for Advanced Study.  This work was supported in part by NSF
grant AST-9820686.

\vspace*{1cm}
\plottwo{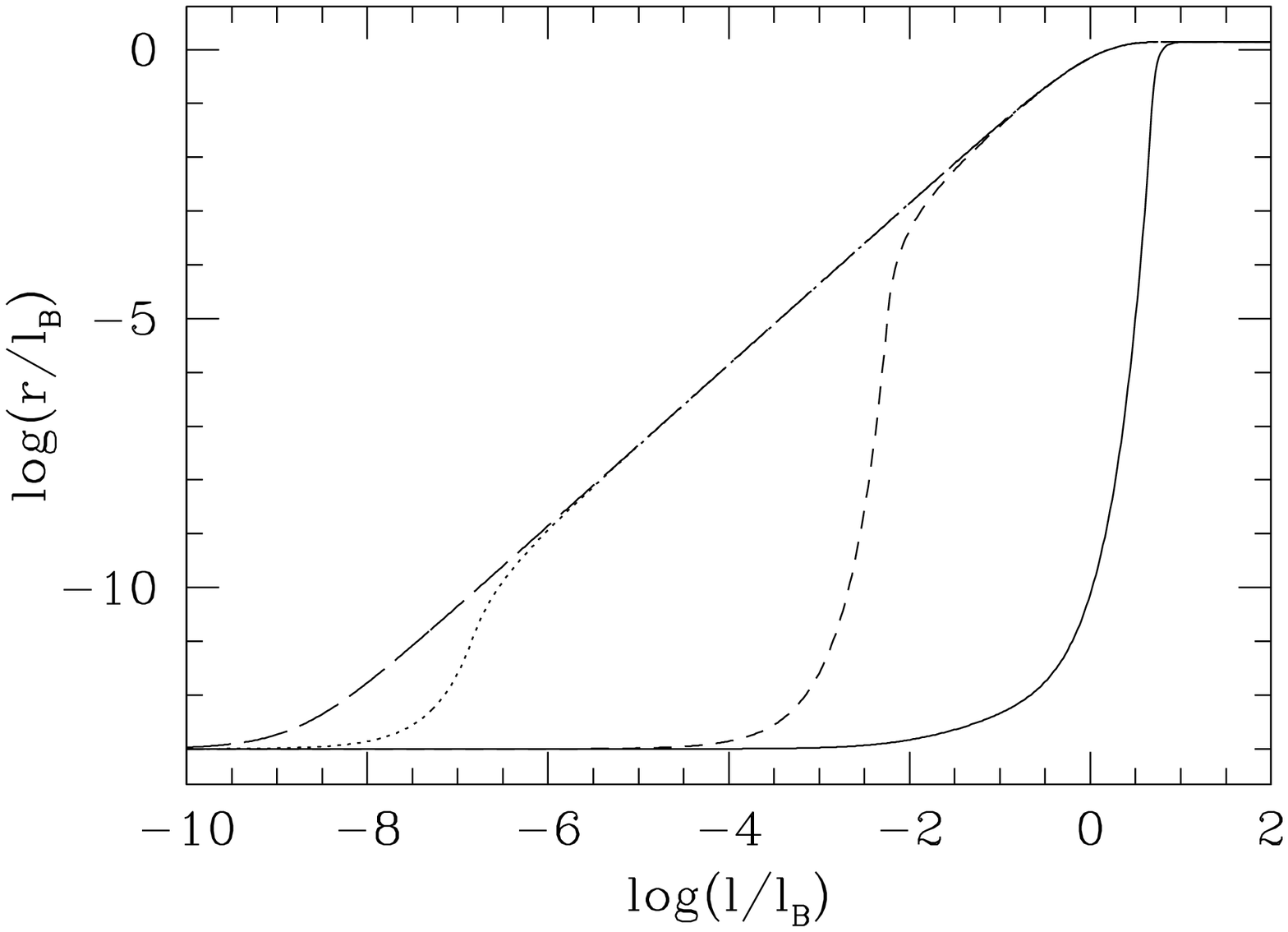}{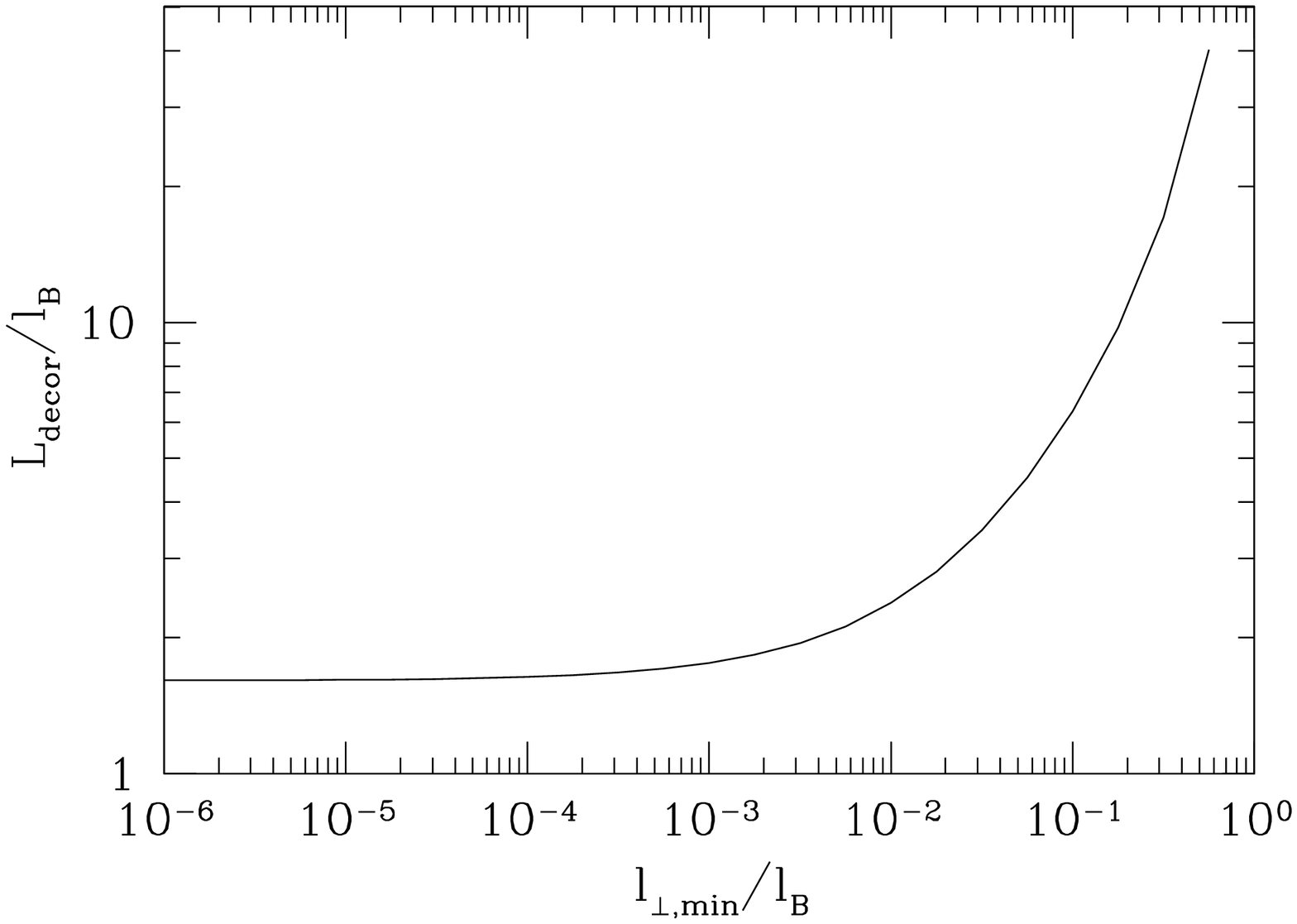}
\vspace*{-2cm}
\figcaption[fig-soln.eps]{Variation of the rms transverse displacement
$r$ of two field lines vs longitudinal distance $l$ along the field
lines, for $\alpha=2/3$.  Four choices of the minimum scale of
turbulence are shown, $\lpm/l_B=10^{-1}$ (solid line), $10^{-5}$
(dashed), $10^{-11}$ ($\sim\rho_p/l_B$, dotted), $10^{-13}$
($\sim\rho_e/l_B$, long-dashed), which correspond to
$l_{\parallel,min}/l_B = 2.2\times10^{-1}, ~4.6\times10^{-4},
~4.6\times10^{-8}, ~2.2\times10^{-9}$, respectively.
\label{f:soln} }

\figcaption[fig-decor.eps]{Variation of the decorrelation length
$L_{\rm dec}$ vs $\lpm$.  Note that $L_{\rm dec}\la2l_B$ for
$\lpm\la10^{-2}l_B$.
\label{f:decor} }

\begin{table}
\caption{Comparison of the Cooling Time and the Conduction Time in
Hydra A and 3C295}
\begin{center}
\begin{tabular}{cccccccc} \hline \hline
\\
Cluster & $R$ & $n$ & $kT$ & $\lambda$ & $t_{\rm cool}$ & $t_{\rm Sp}$ 
& $t_{\rm turb}$ \\
name & (kpc) & (cm$^{-3}$) & (keV) & (kpc) & (Gyr) & (Gyr) & (Gyr) \\
\\
\hline
\\
Hydra A & 100 & 0.005 & 3.6 & 0.8  & 5   & 0.5  & 2 \\
  $''$  & 10  & 0.06  & 3.1 & 0.05 & 0.5 & 0.09 & 0.4 \\
 3C295  & 100 & 0.008 & 5.0 & 0.9  & 7   & 0.3  & 2 \\
  $''$  & 10  & 0.15  & 3.0 & 0.02 & 0.3 & 0.2  & 1 \\
\\
\hline
\end{tabular}
\label{t:1}
\end{center}
\end{table}

\rem{
\newpage
\plotone{fig-soln.eps}
\newline
\plotone{fig-decor.eps}
}
\end{document}